\documentclass[twocolumn,superscriptaddress,preprintnumbers,amsmath,amssymb]{revtex4}
\usepackage{graphicx}
\usepackage{dcolumn}
\usepackage{bm}
\usepackage{color}
\usepackage{dcolumn}

\makeatletter
\@ifundefined{textcolor}{}
{%
 \definecolor{BLACK}{gray}{0}
 \definecolor{WHITE}{gray}{1}
 \definecolor{RED}{rgb}{1,0,0}
 \definecolor{GREEN}{rgb}{0,1,0}
 \definecolor{BLUE}{rgb}{0,0,1}
 \definecolor{CYAN}{cmyk}{1,0,0,0}
 \definecolor{MAGENTA}{cmyk}{0,1,0,0}
 \definecolor{YELLOW}{cmyk}{0,0,1,0}
}

\begin{document}

\title{Spin-orbit-coupled Bose-Einstein condensates of rotating polar molecules}

\author{Y. Deng}
\affiliation{Laboratory of Quantum Engineering and Quantum Metrology, School of Physics and Astronomy, Sun Yat-Sen University (Zhuhai Campus), Zhuhai 519082, China}
\affiliation{State Key Laboratory of Low Dimensional Quantum Physics, Department of Physics, Tsinghua University, Beijing 100084, China}

\author{L. You}
\affiliation{State Key Laboratory of Low Dimensional Quantum Physics, Department of Physics, Tsinghua University, Beijing 100084, China}
\affiliation{Collaborative Innovation Center of Quantum Matter, Beijing 100084, China}

\author{S. Yi}
\affiliation{CAS Key Laboratory of Theoretical Physics, Institute of Theoretical Physics, Chinese Academy of Sciences, Beijing 100190, China}
\affiliation{School of Physics $\&$ CAS Center for Excellence in Topological Quantum Computation, University of Chinese Academy of Sciences, Beijing 100190, China}

\date{\today}

\begin{abstract}
An experimental proposal for realizing spin-orbit (SO) coupling of pseudospin-1 in the ground manifold $^1\Sigma(\upsilon=0)$ of (bosonic) bialkali polar molecules is presented. The three spin components are composed of the ground rotational state and two substates from the first excited rotational level. Using hyperfine resolved Raman processes through two select excited states resonantly coupled by a microwave, an effective coupling between the spin tensor and linear momentum is realized. The properties of Bose-Einstein condensates for such SO-coupled molecules exhibiting dipolar interactions are further explored. In addition to the SO-coupling-induced stripe structures, the singly and doubly quantized vortex phases are found to appear, implicating exciting opportunities for exploring novel quantum physics using SO-coupled rotating polar molecules with dipolar interactions.
\end{abstract}

\maketitle

\section{Introduction}
Recent advances in the experimental realization and manipulation of ultracold polar molecules in the rovibrational ground state~\cite{KRb-exp,LiCs-exp,RbCs-exp,RbCs-exp2,RbCs-exp3,NaK-exp,Guo16,Rvachov17} offer unprecedent scientific opportunities to explore fundamental quantum phenomena and applications, ranging from ultracold chemistry and collisions~\cite{krb-coll, krb-chem,Miranda11} to quantum information processing~\cite{qu-info,qucompu}, simulation of quantum magnetism~\cite{Micheli06,Carr09}, and precise fundamental physics~\cite{pre-measu1,pre-measu2,pre-measu3,Cairncross17}. Of particular interest are spinor polar molecules with internal structures and large electric dipole moments that can be employed to study a host of interesting dipolar effects, such as spontaneous demagnetization~\cite{Pasquiou11}, Fermi surface deformation~\cite{Aikawa14}, and self-bound droplets~\cite{Kadau16,Schmitt16,Barbut16}.

A second area witnessing great progress in atomic quantum gases concerns spin-orbit (SO) coupling for both bosonic~\cite{Lin11, Zhang12, Qu13, Wu15, Campbell16, Luo16, Li17} and fermionic atomic species~\cite{Wang12, Cheuk12, Burdick16, Huang15, Song16}. Spin-orbit-coupled atomic condensates with magnetic dipoles represent a prominent example that combines the advantages of both SO coupling and long range interactions. They are predicted to display interesting quantum phases~\cite{Deng2012,Wilson2013,Ng2014}. Their counterparts, electric dipolar condensates of molecules with coupling between the rotational and orbital angular momenta~\cite{Deng2015,Wall15}, present an equally promising platform if a SO-like coupling can be identified. Although hyperfine resolved two-photon transfer~\cite{NaK-exp,Guo16} has been realized experimentally, an analogous atomic SO interaction cannot be directly engineered this way because the neighboring rotational levels for rotating molecules possess opposite parities. This experimental challenge for realizing SO coupling with two-photon Raman processes in dipolar quantum gases of molecules has not been thoroughly investigated yet. Whether it can be overcome in realistic spinor condensates of rotating polar molecules or not is the question to be addressed in this study.

\begin{figure}[tbp]
\includegraphics[width=0.8\columnwidth]{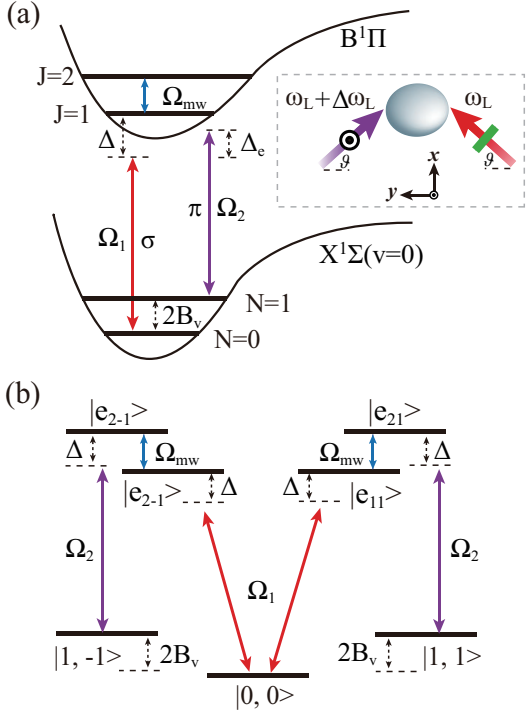}
\caption{(color online). (a) Schematic for creating SO coupling of a polar molecule. The linearly polarized $\sigma$-($\pi$-) laser field 1(2) drives the hyperfine resolved single-photon transition between $X^1\Sigma^+ (v=0)$ rotational ground states and $B^1\Pi$ vibrational excited states. A $\pi$-polarized microwave field resonantly connects two excited states with opposite parities. (b) Relevant level structure of a polar molecules. Here $|1,\pm1\rangle$ ($|0,0\rangle$) denotes the hyperfine level of the $N=1$ ($N=0$) rotational state and $|e_{1\pm1}\rangle$ ($|e_{2\pm1}\rangle$) denotes the Zeeman level of the electronically excited $J=1$ ($J=2$) state.}
\label{model}
\end{figure}

In this work we propose the synthesis of SO coupling in an ultracold Bose gas of polar molecules. The pseudospin $1$ consists of the ground rotational level and two substates of the first excited rotational level. Their SO coupling is created through hyperfine resolved Raman processes through two excited states of opposite parities resonantly coupled by a microwave. The laser configuration ensures that the electric dipole-dipole interaction (DDI) between the rotational and orbital degrees of freedom persists, and as a result the synthetic SO coupling facilitates many interesting quantum phases. A doubly quantized vortex phase surprisingly emerges in the spin state with the highest occupation.

This paper is organized as follows. In Sec.~\ref{model} we introduce our scheme for generating SO coupling and derive the single-particle Hamiltonian for a pseudospin-1 molecule. In Sec.~\ref{minte} we derive the contact $s$-wave interactions and the effective DDI between two pseudospin-1 molecules. The quantum phases of the SO-coupled spinor condensates of rotating polar molecules are presented in Sec.~\ref{results}. A brief summary is given in Sec.~\ref{cons}.

\section{Model}\label{model}
We consider a gas of ultracold bialkali polar molecules in the rovibrational ground state, $X{}^1\Sigma(v=0)$, characterized by the rotational angular momentum ${\mathbf N}$ and nuclear spins ${\mathbf I}_{i}$ ($i=1,2$). The internal state of the molecule is thus denoted by $|M_{1}M_{2}NM_{N}\rangle$, where $M_{i}$ and $M_{N}$ are, respectively, the projections of ${\mathbf I}_{i}$ and ${\mathbf N}$ along the quantization axis. As shown in Ref.~\cite{Deng2015}, under a strong bias magnetic field, e.g., ${\bf B}=B\hat{z}$, $M_{i}$ becomes a good quantum number, which can be fixed for instance, by choosing $M_{i}=I_{i}$. Therefore, the relevant internal states reduce to $|N,M_N\rangle=|0,0\rangle$, $|1,0\rangle$, and $|1,\pm1\rangle$ at sufficiently low temperatures. The energy gap between the ground state $N=0$ and the first excited state $N=1$ is $2B_{v}$, with $B_{v}$ ($\sim10\,$GHz) being the rotation constant. Among the three states in the $N=1$ manifold, it is shown quite generally that the states $|1,1\rangle$ and $|1,-1\rangle$ can be tuned near degeneracy and well separated from $|1,0\rangle$~\cite{Deng2015}. Consequently, we may ignore $|1,0\rangle$ and focus on the three internal states $|0,0\rangle$ and $|1,\pm1\rangle$. Their quantum numbers $M_N$ serve as shorthand notation for a pseudospin $1$.

The SO coupling involving the ground state pseudospin $1$ is created via Raman processes to an electronically excited intermediate level, e.g., ${B}^1\Pi$ (Fig.~\ref{model}), whose rotational states $J=1$ and $2$ are split by an energy gap $\hbar \Delta_e$. Here $\mathbf{J}$ represents the total angular momentum excluding the nuclear spins. Limited by the parity and the electric dipole transition selection rules, the ground state $|N=0\rangle$ ($|N=1\rangle$) can only couple to the excited $|J=1\rangle$ ($|J=2\rangle$) state of opposite parity via single photon transitions. Therefore, to effect Raman transitions, the two excited states are mixed by a position-independent $\pi$-polarized microwave field with Rabi frequency $\Omega_{\rm{mw}}$. As illustrated in Fig.~\ref{model}(a), two linearly polarized plane-wave lasers drive, respectively, the molecular transitions $|N=0\rangle\leftrightarrow|J=1\rangle$ and $|N=1\rangle\leftrightarrow|J=2\rangle$ with Rabi frequencies $\Omega_{1}e^{i{\bf k_{1}\cdot r}}$ and $\Omega_{2}e^{i{\bf k_{2}\cdot r}}$. Here ${\mathbf k}_{1,2}=k_{L}(\sin\vartheta \hat {\mathbf x} \pm\cos\vartheta \hat {\mathbf y})$ are the laser wave vectors, and the angle $\vartheta$ is tunable. The frequencies of the lasers 1 and 2 are $\omega_{L}$ and $\omega_{L}+\Delta\omega_{L}$, respectively, with $|\Delta\omega_{L}|\approx |\Delta_e-2B_{\upsilon}/\hbar|$ ($\ll\omega_{L}$) chosen to compensate for the energy splitting between the $N=0$ and $1$ rotational states. The frequency $\omega_L$ is detuned from the resonance frequency by $\Delta$ and operates in the limit of large detuning $|\Omega_{1,2}/\Delta|\ll 1$. Adiabatically eliminating the two excited states, the single-molecule Hamiltonian becomes (see the Appendix)
\begin{align}
\hat h = \begin{pmatrix}
\frac{{\mathbf p}^2}{2M}+ \hbar\delta_1+\hbar\delta_2& \hbar\Omega e^{i\kappa y} & 0 \\
\hbar\Omega  e^{-i\kappa y}  & \frac{{\mathbf p}^2}{2M}  & \hbar\Omega e^{-i\kappa y}\\
0 &\hbar\Omega e^{i\kappa y}  & \frac{{\mathbf p}^2}{2M}-\hbar\delta_1 +\hbar\delta_2
 \end{pmatrix}\label{sing}
\end{align}
in the basis $\{|1\rangle,|0\rangle,|-1\rangle\}$, where $\mathbf{p}$ denotes the momentum, and $M$ is the molecular mass. In addition, $\delta_1$ and $\delta_{2}$ are independently tunable (see the Appendix), representing the effective linear and quadratic Zeeman shifts, respectively,  $\Omega={\Omega_{\rm{mw}}\Omega_1\Omega_2}/({\Delta^2 -\Omega_{\rm{mw}}^2})$ is the Raman coupling strength, and $\kappa=2k_{L}\cos\vartheta$, which can also be tuned independently.

After applying the transformation $|0\rangle \rightarrow |0\rangle e^{i\kappa y/2}$ and $|\pm1\rangle \rightarrow |\pm1\rangle e^{-i\kappa y/2}$,in the $y$ direction, the Hamiltonian~(\ref{sing}) becomes
\begin{align}
\hat{h}'_y&=\frac{\hbar^2\left(q^2 - \kappa q\right)}{2M}\hat{I} +\frac{\hbar^2\kappa}{M}q\hat{S}_z^2\nonumber\\
&\quad + \sqrt{2}\hbar\Omega\hat{S}_x +\hbar\delta_1\hat{S}_z+\hbar\delta_2\hat{S}_z^2,\label{sing2}
\end{align}
where $q={p}_y/\hbar$ defines the quasimomentum, $\hat{I}$ is the identity matrix, and $\hat{S}_{x,y,z}$ are the Pauli matrices for spin-1 particles. Unlike the nominal SO-coupling term $q \hat{S}_z$ already discussed in Raman-dressed spin-1 atoms~\cite{Lan2014,Natu2015}, the SO coupling we realize here is of the form $q\hat{S}_z^2$, acting as an effective momentum-dependent quadratic Zeeman shift, a coupling between the spin tensor and linear momentum. This same term was recently proposed by Luo {\it et al}. in a spin-1 atom by introducing an extra Raman laser~\cite{Luo17}. The competition of the spin-tensor-momentum coupling and the short-range spin-exchange interaction is shown to cause a different type of striped superfluid.

\begin{figure}[tbp]
\includegraphics[width=0.95\columnwidth]{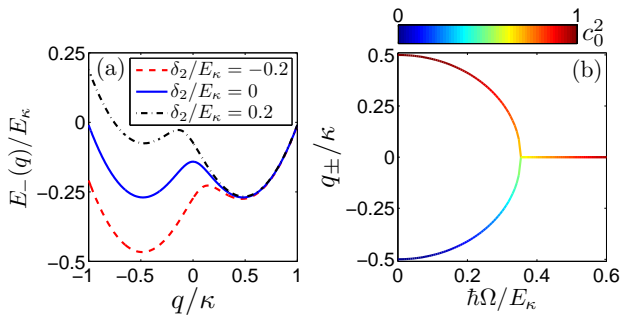}
\caption{(color online). (a) Lower branch dispersion for different values of $\delta_2$ with $\Omega/E_{\kappa}=0.1$ and $\delta_1=0$. (b) Local minima of quasimomentum $q_{\pm}$ as a function of $\Omega$ for $\delta_1=\delta_2=0$, with the blue to red color gradient calibrating the population of the $|0\rangle$ state. The ground-state branch displays a left edge corresponding to the superposition of $|\pm1\rangle$ states and a right edge corresponding to the $|0\rangle$ state.}
\label{single-E}
\end{figure}

Take the simple case of $\delta_1=0$, diagonalizing the Hamiltonian (\ref{sing2}). The eigensystem reveals two bright states $E_{\pm}(q) = \frac{{\hbar^{2} q^2}}{2M} +\frac{\hbar\delta_2}{2}\pm \hbar{\sqrt{\big(\frac{\hbar{\kappa} q}{2M} +\frac{\delta_2}{2} \big)^2 + 2\Omega^2}}$ and one dark state $E_{0}(q) = \frac{\hbar^{2}(q^2+\kappa q)}{2M}+{\hbar\delta_2}$. Among these three, the $E_{+}(q)$ branch is the highest in energy and can be left out of the discussion on low-energy physics. Moreover, independent of $q$, the eigenstate corresponding to $E_{0}(q)$ always takes the form $|\zeta\rangle\equiv(|1\rangle-|-1\rangle)/\sqrt{2}$ with the band minimum located at $q/\kappa=-1/2$. The $E_{0}(q)$ branch does not possess SO coupling and remains orthogonal to the bright-state branch $E_{-}(q)$. As a result, we only need to focus on the $E_-(q)$ branch for the single-particle spectrum.

Figure~\ref{single-E}(a) shows the dependence of $E_-(q)$ on $\delta_2$ for $\Omega/E_{\kappa}=0.1$, where $E_\kappa\equiv\hbar^2\kappa^2/2M$ is adopted as the energy unit. The dispersion curve $E_-(q)$ displays the characteristic double well structure as in atomic SO coupling in the quasi-momentum space. The eigenstates corresponding respectively to the local minima of the left and right wells are dominated by the spin states $|\zeta\rangle$ and $|0\rangle$. Consequently, the energy of the left (right) local minimum is sensitive (immune) to the quadratic Zeeman parameter $\delta_2$. At sufficiently large $|\delta_2|$, the double-well dispersion becomes a single well such that only the left (right) well remains when $\delta_2<0$ ($\delta_2>0$). To study how $E_-(q)$ depends on the Rabi coupling strength $\Omega$, we consider the simplest case with $\delta_2=0$. It is then easy to show that $E_-(q)$ possesses two local minima at $q_{\pm} = \pm \kappa \sqrt{1 - 8(\hbar\Omega/E_\kappa)^2}/2$ when $\hbar\Omega <\sqrt{2} E_{\kappa}/4$ and a single minimum at $q=0$ when $\hbar \Omega >\sqrt{2} E_{\kappa}/4$. Figure~\ref{single-E}(b) plots the $\Omega$ dependence of $q_\pm$ for $\delta_1=\delta_2=0$. In this case, the eigenstate wave functions at these minima can be generally expressed as $c_\zeta|\zeta\rangle+c_0|0\rangle$. The $\Omega$ dependence of $|c_0|^2$ is as shown in Fig.~\ref{single-E}(b).

\section{Molecular Interactions}\label{minte}
We next include interactions between molecules. We note that the condensate consisting of molecules in the $N=0$ and $1$ rotational levels. The collisions between two spin-1 molecules are characterized by the scattering lengths $a^{(11)}_{0}$ and $a^{(11)}_{2}$ respectively corresponding to the collisional channels with total rotational angular momenta $N_{\rm tot}=0$ and $2$. The interaction between distinguishable rotational levels $N=0$ and $1$ molecules is described by the scattering length $a^{(01)}$. Finally, the collision between two spin-0 molecules is characterized by $a^{(00)}$~\cite{youxu}. In the reduced Hilbert space in the lowest energy branch, the Hamiltonian for the contact interactions becomes
\begin{align}
\hat{\cal H}_{\rm ci}&=\int d{\mathbf
r}:\bigg[\frac{g^{(00)}}{2}\hat n_{0}^{2}+\frac{g_{2}^{(11)}}{2}\left(\hat n_{1}^{2}+\hat n_{-1}^{2}\right)\nonumber\\
&\quad+g^{(01)}\left(\hat n_{1}\hat n_{0}+\hat n_{0}\hat n_{-1}\right)+g^{(11)}\hat n_{1}\hat n_{-1}\bigg]:\,,\label{hcol}
\end{align}
where $\hat n_{\sigma}({\mathbf r})=\hat \psi_{\sigma}^{\dag}({\mathbf r})\hat\psi_{\sigma}({\mathbf r})$ with $\hat{\psi}_{\sigma}({\mathbf r})$ being the field operators for spin-$\sigma$ ($\sigma=0,\pm1$) molecules, $g^{(00)}=4\pi\hbar^{2}a^{(00)}/M$, $g^{(01)}=4\pi\hbar^{2}a^{(01)}/M$, $g_{2}^{(11)}\!=\!4\pi\hbar^{2}a_{2}^{(11)}/M$, $g^{(11)}=4\pi\hbar^{2}(2a_{0}^{(11)}+a_{2}^{(11)})/3M$, and $:\hat O:$ arranges operator in normal order. Unlike spin-1 atomic gases, for polar molecules the contact spin-exchange interaction is absent from $\hat {\mathcal H}_{\rm ci}$
because of the pseudospin-1 construct.

The DDI between polar molecules is treated as in Ref.~\cite{Deng2015}, which takes the following simplified form in the reduced Hilbert space:
\begin{align}
\hat {\cal H}_{\rm dd}&=
{g_d}\sqrt{\frac{4\pi}{45}}\int \frac{d {\bf r} d {\bf
r'}}{|{\mathbf r}-{\mathbf
r}'|^3} \Bigg\{\left[\sqrt{6}Y_{22}({\mathbf
e})\hat{\psi}_{0}^\dag\hat{\psi}_{-1}^{'\dag}\hat{\psi}'_{0} \hat{\psi}_{1} \!+\! {\rm H.c.}\right] \nonumber \\
 &\quad + Y_{20}({\mathbf
e})\left[\hat{\psi}_{0}^\dag\hat{\psi}_{1}^{'\dag}\hat{\psi}'_{0} \hat{\psi}_{1}
+\hat{\psi}_{0}^\dag\hat{\psi}_{-1}^{'\dag}\hat{\psi}'_{0}\hat{\psi}_{-1}\right] \Bigg\}.
\label{smhdd}
\end{align}
In the above, $g_{d}=d^{2}/4\pi\epsilon_{0}$ characterizes the strength of the DDI with $d$ being the molecule permanent dipole moment and $\epsilon_{0}$ the electric permittivity of vacuum, ${\mathbf{e}} = ({\mathbf {r}} - {\mathbf {r}'})/|{\mathbf {r}} - {\mathbf {r}'}|$ is a unit vector, and $Y_{20}({\mathbf e})$ and $Y_{22}({\mathbf e})$ are spherical harmonics. We use the shorthand notation $\hat{\psi}_{\sigma}\equiv\hat{\psi}_{\sigma}({\mathbf r})$ and $\hat{\psi}'_{\sigma}\equiv\hat{\psi}_{\sigma}({\mathbf r'})$. The DDI $\hat H_{\rm dd}$ couples rotational and orbital angular momenta, within the constraint that the total angular momentum is conserved. Under the analogous configuration of Ref.~\cite{Luo17} for atomic spin-1 gases, the SO coupling in the magnetic DDI, however, is suppressed by the bias magnetic field~\cite{Ueda2007}.

The quantum phases for the SO-coupled molecular condensate can be found with a mean-field treatment which replaces the field operator $\hat\psi_{\sigma}$ by its mean value $\psi_{\sigma}\equiv\langle \hat\psi_{\sigma}\rangle$. Here $\psi_{\sigma}$ is found numerically by minimizing the energy functional ${\cal F}[\psi_{\sigma},\psi_{\sigma}^{*}]=\langle\hat{\cal H}_{0}+\hat{\cal H}_{\rm ci}+\hat{\cal H}_{\rm dd}\rangle$, where
\begin{eqnarray}
\hat{\cal H}_{0}=\sum_{\sigma\sigma'}\int d{\mathbf r}\hat\psi_{\sigma}^{\dag}({\mathbf r})\left[\hat h_{\sigma\sigma'}+U({\mathbf r})\delta_{\sigma\sigma'}\right]\hat\psi_{\sigma}({\mathbf r}),
\end{eqnarray}
with $U({\mathbf r})= M\nu_{\perp}^2(x^2+y^2+\gamma^2z^2)/2$ an axially symmetric trap of radial frequency $\nu_\perp$ and trap aspect ratio $\gamma$. To proceed further, we consider, without loss of generality, a condensate of ${\mathcal {N}}=5\times 10^4$ LiNa molecules with $d=0.45\,$D. The external trap parameters are chosen to be $\nu_\perp=(2\pi)100\,$Hz and $\gamma=6.3$ (oblate). For simplicity, we treat the condensate as a quasi-two-dimensional (quasi-2D) one by decomposing the condensate wave functions into $\psi_{\sigma} ({\bf r}) = \phi_{\sigma}(x,y)\phi_z(z)$, where $\phi_{\sigma}(x,y)$ is the two-dimensional wave function and $\phi_z(z) = (\gamma/\pi \ell_{\perp}^2)^{1/4} e^{-\gamma z^2/2 \ell_{\perp}^2}$ is the ground state of the axial harmonic trap with $\ell_{\perp}=\sqrt{\hbar/M\omega_{\perp}}$. We assume that the Rabi frequency is $\Omega=0.07 E_{\kappa}$, with $E_{\kappa}/\hbar=(2\pi)\,1.7\,$kHz, which consequently fixes the SO-coupling strength $\kappa$. We further assume that all $s$-wave scattering lengths are equal to $150\,a_{B}$ with $a_{B}$ being the Bohr radius. Thus, the free parameters in our model reduce to the linear Zeeman shift $\delta_1$ and quadratic Zeeman shift $\delta_2$.

\begin{figure}[tbp]
\includegraphics[width=0.95\columnwidth]{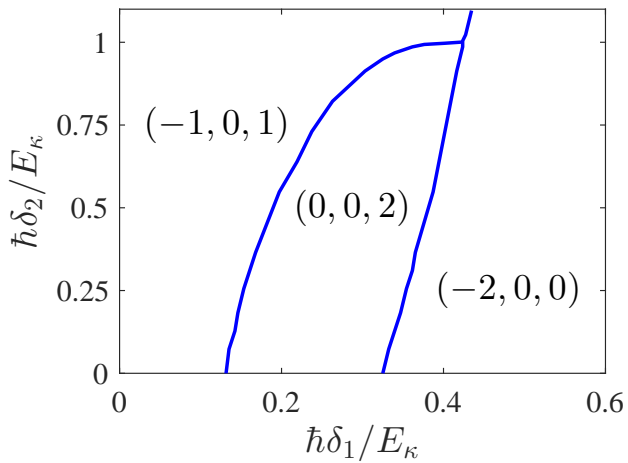}
\caption{(color online). Mean field phase diagram in the $\delta_1$-$\delta_2$ parameter plane.} \label{manyphase}
\end{figure}

\section{Results}\label{results}
\begin{figure}[tbp]
\includegraphics[width=0.95\columnwidth]{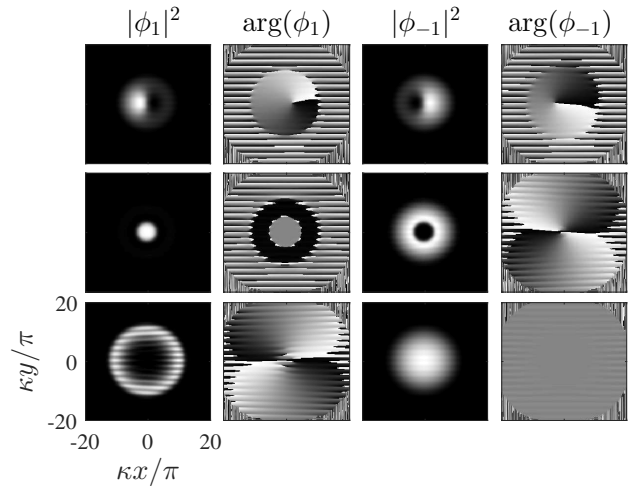}
\caption{(color online). Typical condensate wave functions for the $(-1,0,1)$ phase (row 1), $(0,0,2)$ phase (row 2), and $(-2,0,0)$ phase (row 3). Columns 1 and 3 represent the densities $|\phi_{1}|^2$ and  $|\phi_{-1}|^2$, respectively, while columns 2 and 4 plot their respective phases $\arg({\phi_{1}})$ and  $\arg({\phi_{-1}})$. The other parameters are $\hbar\delta_2/E_{\kappa}=0.55$ and, from row 1 to row 3, $\hbar\delta_1/E_{\kappa}=0.07$, $0.29$, and $0.40$, respectively.} \label{wavefunc}
\end{figure}

Figure~\ref{manyphase} summarizes the phase diagram in the $\delta_1$-$\delta_2$ parameter plane where each phase is denoted by the winding number configuration for the corresponding wave functions, i.e., $(w_1,w_0,w_{-1})$. In Fig.~\ref{wavefunc} we plot the typical wave functions $\phi_1$ and $\phi_{-1}$ for several different phases; $\phi_0$ is not shown because it does not contain any interesting structure. In the $(-1,0,1)$ phase, $\phi_{1}$ and $\phi_{-1}$ exhibit singly quantized vortices with opposite winding numbers, while for the $(0,0,2)$ [$(-2,0,0)$] phase, only $\phi_{-1}$ ($\phi_1$) displays a doubly quantized vortex induced by the $Y_{2,\pm2}$ terms in the DDI [Eq.~\eqref{smhdd}]. To understand this, we consider, for example, the $Y_{22}$ term. The process of transferring a molecule in state $|1\rangle$ to state $|-1\rangle$ lowers the spin angular momentum projection by $2\hbar$. To conserve the total angular momentum, the orbital angular momentum associated with the molecule spin state $|-1\rangle$ must be increased by $2\hbar$. Hence the winding numbers for the wave functions in different phases satisfy $w_{-1}-w_1=2$. The stripes in the density and phase plots of the wave functions are due to the interplay between the SO coupling and the contact interactions. They are not the focus of the present work since we have taken the simple-minded approach of a constant $s$-wave scattering length for all.

\begin{figure}[tbp]
\includegraphics[width=0.96\columnwidth]{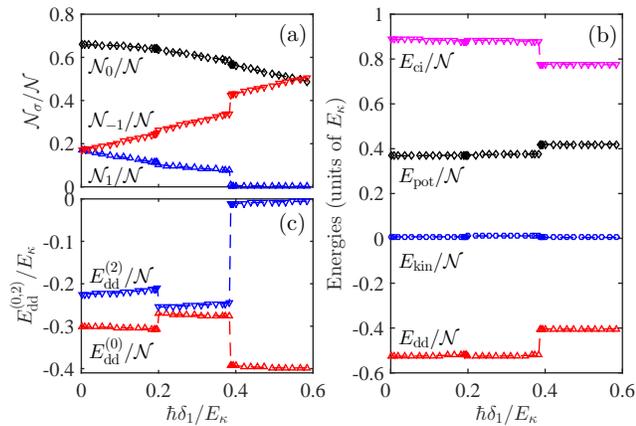}
\caption{(color online). The $\delta_1$ dependence for various quantities at $\hbar\delta_2/E_{\kappa}=0.55$: (a) occupation numbers $\mathcal{N}_\sigma$; (b) energy components $E_{\rm ci}$, $E_{\rm pot}$, $E_{\rm kin}$, and $E_{\rm dd}$; and (c) DDI energies $E^{(0)}_{\mathrm{dd}}$ and $E^{(2)}_{\mathrm{dd}}$.}\label{OPdel2}
\end{figure}

To gain more insight into these quantum phases, we plot, in Fig.~\ref{OPdel2}(a), the $\delta_1$ dependence of the molecule number ${\cal N}_{\sigma}=\int dxdy|\phi_{\sigma}|^{2}$ for $\hbar\delta_2/E_{\kappa}=0.55$. As can be seen, for a given $\delta_2$, the spin $|0\rangle$ state generally has the highest occupation number unless $\delta_1$ becomes very large. In particular at $\delta_1=0$, we have $\mathcal{N}_{1}=\mathcal{N}_{-1}$, hence it is energetically favorable if both $\phi_1$ and $\phi_{-1}$ are singly quantized vortices,i.e., the $(-1,0,1)$ phase, since its kinetic energy is lower than the multiply quantized vortex state. For small $\delta_1$, the condensate remains in this quantum phase even when $\mathcal{N}_{-1}$ is slightly larger than $\mathcal{N}_{1}$. When $\delta_1$ becomes sufficiently large, $\mathcal{N}_{-1}$ is significantly larger than $\mathcal{N}_{1}$. The highly populated $\phi_{-1}$ is vortex-free, while the less populated $\phi_{1}$ state becomes a doubly-quantized vortex, i.e., in the $(-2,0,0)$ phase, in order to lower the kinetic energy associated with the vortices. Surprisingly, the phase diagram in Fig.~\ref{manyphase} also reveals an unusual region of $(0,0,2)$ phase if $\hbar\delta_2/E_{\kappa}\lesssim 1.0$, where, as shown in Fig.~\ref{OPdel2}(a), $\mathcal{N}_{-1}$ is notably larger than $\mathcal{N}_{1}$. To understand this, we plot in Fig.~\ref{OPdel2}(b) the $\delta_1$ dependence of the kinetic energy $E_{\rm kin}$, the potential energy $E_{\rm pot}$, the contact interaction energy $E_{\rm ci}$, and the DDI energy $E_{\rm dd}$. In contrast to the large change of the DDI energy across the $(0,0,2)$-to-$(-2,0,0)$ transition, the kinetic energy in the $(0,0,2)$ phase is only slightly larger than in the two other phases. The kinetic energy is thus not the main force that drives the phase transitions. Instead, the dipolar interaction energy deserves a closer analysis. Figure~\ref{OPdel2}(c) plots the $\delta_1$ dependence of $E_{\rm dd}^{(0)}$ and $E_{\rm dd}^{(2)}$, the energy components of the DDI originated from the $Y_{20}$ and $Y_{2\pm2}$ terms, respectively. As can be seen, both terms give rise to negative interaction energies except that, in the $(-2,0,0)$ phase, the $Y_{2\pm2}$ contributions are negligibly small due to the nearly vanishing $\mathcal{N}_1$. While in the $(0,0,2)$ phase, the contributions from the $Y_{2\pm2}$ terms become significant since the occupation numbers of all spin states are comparable. Finally, we explain the $\delta_2$ dependence of the phase diagram. By increasing $\delta_2$, the occupation numbers $\mathcal{N}_1$ and $\mathcal{N}_{-1}$ both drop such that the DDI originating from the $Y_{2\pm2}$ terms are suppressed. Consequently, the $(0,0,2)$ phase disappears for sufficiently large $\delta_2$.

\section{Conclusions}\label{cons}
We proposed a scheme to realize spin-tensor-momentum coupling in a gas of rotating polar molecules by combining the hyperfine resolved Raman processes with the microwave field coupled rotational states in the first electronic excited manifold. Under suitable conditions, we showed that the electric DDI remains effective in coupling the rotation and orbit degrees of freedom of the molecules. We further explored the ground-state properties of the SO-coupled molecular condensate with dipolar interactions and mapped out its zero-temperature phase diagram. The interplay between the linear Zeeman shift and the DDI gives rise to the singly and doubly quantized vortex phases, while the spin-tensor-momentum coupling imprints stripes on the condensate wave functions. The proposed scheme seems within reach of the leading effects on ultracold polar molecule experiments. It opens several interesting opportunities for exploring interesting physics with molecular quantum gases, such as spin vortex matter in superfluidity~\cite{Sonin87,Kopnin02,Bloch08,Fetter09}, droplets with quantum fluctuations~\cite{Kadau16,Schmitt16,Barbut16}, strongly correlated many-body physics~\cite{atom-dipspin-theo1,Yi2007,atom-dipspin-exp1}, and exotic topological quantum phases~\cite{Yao2012,Yao2013,Peter2015,Syzranov2015}.

\section{ACKNOWLEDGMENT}\label{acknow}
This work was supported by the NSFC (Grants No. 11434011, No. 11674334, No. 11654001, and No. 11747605) and by NKRDPC (Grant No. 2017YFA0304501).

\appendix
\begin{widetext}

\section{Single-particle Hamiltonian}
In this appendix we present the details on the derivation of the  single-molecule Hamiltonian for realizing spin-tensor-momentum coupling of pseudospin-1 bialkali polar molecules. For the given level diagram and laser configurations in Fig.~\ref{model}(b) of the main text, the single-molecule Hamiltonian in the Raman fields reads
\begin{eqnarray}
\hat{\cal H}  &=&  2B_{v}\sum_{\sigma=\pm1}|1,\sigma\rangle\langle 1,\sigma| +\hbar\delta_{1,-1}|1,1\rangle\langle
1,1|+\hbar \omega_e \sum_{\sigma=\pm1}|e_{1\sigma}\rangle\langle e_{1\sigma}|+\hbar (\omega_e+\Delta_e) \sum_{\sigma=\pm1}|e_{2\sigma}\rangle\langle e_{2\sigma}| \nonumber \\
&+&\hbar\sum_{\sigma=\pm1}\left(\Omega_{\rm{mw}}e^{-i\Delta_{e}t}|e_{2\sigma}\rangle\langle e_{1\sigma}|+\Omega_1({\bf r})e^{-i\omega_{L}t}|e_{1\sigma}\rangle\langle 0,0|+\Omega_2({\bf r})e^{-i(\omega_{L}+\Delta \omega_L)t}|e_{2\sigma}\rangle\langle 1,\sigma|+{\rm H.c.}\right),
\end{eqnarray}
where $|e_{1\sigma}\rangle$ ($|e_{2\sigma}\rangle$) is the Zeeman level of the electronically excited state for $J=1$ ($J=2$), $\hbar \omega_e$ is the transition frequency of $|N=0\rangle\leftrightarrow|J=1\rangle$, and $\delta_{1,-1}$ is the hyperfine splitting for single molecules. In the rotating frame, a unitary transformation is introduced as ${\cal U} = \exp(-i\hat{\cal H}'t/\hbar)$ with
\begin{align}
\hat{\cal H}' &=\hbar(\Delta_e-\Delta\omega_L)\sum_{\sigma=\pm1}|1,\sigma\rangle\langle 1,\sigma| + \hbar\omega_L \sum_{\sigma=\pm1}|e_{1\sigma}\rangle\langle e_{1\sigma}|  + \hbar(\omega_L+\Delta_e)\sum_{\sigma=\pm1} |e_{2\sigma}\rangle\langle e_{2\sigma}|.
\end{align}
After applying the unitary transformation, the time-independent Hamiltonian is given by
\begin{align}
\hat{\cal H} &\;\rightarrow {\cal U}^\dag\hat{\cal H}{\cal U} - i\hbar
{\cal U}^\dag\frac{\partial}{\partial t}{\cal U} \nonumber \\
& = \hbar(\delta +
\delta_{1,-1})|1,1\rangle\langle
1,1|+\hbar\delta|1,-1\rangle\langle 1,-1|  +\hbar\Delta\sum_{\zeta=1,2}\sum_{\sigma=\pm1}|e_{\zeta q}\rangle\langle
e_{\zeta q}|+\hbar\sum_{\sigma=\pm1}\Big\{\Omega_{\rm{mw}}|e_{2\sigma}\rangle\langle e_{1\sigma}|\nonumber \\
& +\Omega_{1}({\bf r})|e_{1 q}\rangle\langle 0,0|+\Omega_{2}({\bf r})|e_{2\sigma}\rangle\langle 1,\sigma|+{\rm H.c.}\Big\}, \label{single}
\end{align}
with $\Delta=\omega_{e} -\omega_L $ the molecule-light detuning and $\delta=2B_{\upsilon}/\hbar +\Delta\omega_L-\Delta_e$ the two-photon detuning.

To proceed further, we rewrite the Hamiltonian (\ref{single}) in the second quantized as (without the center-of-mass motion)
\begin{align}
\hat{\cal H} & = \hbar(\delta +
\delta_{1,-1})\hat{\psi}^\dag_{11}\hat{\psi}_{11}+\hbar\delta\hat{\psi}^\dag_{1-1}\hat{\psi}_{1-1} +\hbar\Delta\sum_{\zeta=1,2}\sum_{\sigma=\pm1}\hat{\psi}^\dag_{e_{\zeta q }}\hat{\psi}_{e_{\zeta q}} +\hbar\sum_{\sigma=\pm1}\left[\Omega_{\rm{mw}}\hat{\psi}^\dag_{e_{2\sigma}}\hat{\psi}_{e_{1\sigma}} \right.\nonumber \\
&\left.+\Omega_1({\bf r})\hat{\psi}^\dag_{e_{1\sigma}}\hat{\psi}_{00}+\Omega_2({\bf r})\hat{\psi}^\dag_{e_{2\sigma}}\hat{\psi}_{1\sigma}+{\rm H.c.}\right], \label{many}
\end{align}
where $\hat{\psi}_{NM_{N}}$ is the annihilation field operator for the ground rotational state and $\hat{\psi}_{e_{\zeta q}}$ is the annihilation field operators for the excited rotational state. Then it is straightforward to calculate the equations of motion for the field operators
\begin{align}
i\dot{\hat \psi}_{00} &=  \Omega^*_1({\bf r})\sum_{\sigma=\pm1}\hat{\psi}_{e_{1\sigma}}, \nonumber \\
i\dot{\hat \psi}_{11} &=  (\delta+\delta_{1,-1}){\hat \psi}_{11} +\Omega^*_2({\bf r})\hat{\psi}_{e_{21}} , \nonumber \\
i\dot{\hat \psi}_{1-1} &=  \delta{\hat \psi}_{1-1} +\Omega^*_2({\bf r})\hat{\psi}_{e_{2-1}} , \nonumber \\
i\dot{\hat \psi}_{e_{11}} &=  \Delta{\hat \psi}_{e_{11}} +\Omega_{\rm{mw}}\hat{\psi}_{e_{21}} +\Omega_1({\bf r})\hat{\psi}_{00} , \nonumber \\
i\dot{\hat \psi}_{e_{21}} &=  \Delta{\hat \psi}_{e_{21}} +\Omega_{\rm{mw}}\hat{\psi}_{e_{11}} +\Omega_2({\bf r})\hat{\psi}_{11}, \nonumber \\
i\dot{\hat \psi}_{e_{1-1}} &=  \Delta{\hat \psi}_{e_{1-1}} +\Omega_{\rm{mw}}\hat{\psi}_{e_{2-1}} +\Omega_1({\bf r})\hat{\psi}_{00} , \nonumber \\
i\dot{\hat \psi}_{e_{2-1}} &=  \Delta{\hat \psi}_{e_{2-1}} +\Omega_{\rm{mw}}\hat{\psi}_{e_{1-1}} +\Omega_2({\bf r})\hat{\psi}_{1-1}. \label{dequs}
\end{align}

Under the conditions $|\Omega_{\rm{mw}}/\Delta|\ll1$ and $|\Omega_{1,2}/\Delta|\ll1$, the excited states can be adiabatically eliminated to yield
\begin{align}
\hat\psi_{e_{11}} &=\frac{\Omega_{\rm{mw}}\Omega_2({\bf r})} {\Delta^2 -\Omega_{\rm{mw}}^2}\hat{\psi}_{11} - \frac{\Delta\Omega_1({\bf r})} {\Delta^2 -\Omega_{\rm{mw}}^2}\hat{\psi}_{00}, \nonumber \\
\hat\psi_{e_{21}} &=\frac{\Omega_{\rm{mw}}\Omega_1({\bf r})} {\Delta^2 -\Omega_{\rm{mw}}^2}\hat{\psi}_{00} - \frac{\Delta\Omega_2({\bf r})} {\Delta^2 -\Omega_{\rm{mw}}^2}\hat{\psi}_{11},\nonumber \\
\hat\psi_{e_{1-1}} &=\frac{\Omega_{\rm{mw}}\Omega_2({\bf r})} {\Delta^2 -\Omega_{\rm{mw}}^2}\hat{\psi}_{1-1} - \frac{\Delta\Omega_1({\bf r})} {\Delta^2 -\Omega_{\rm{mw}}^2}\hat{\psi}_{00}, \nonumber \\
\hat\psi_{e_{2-1}} &=\frac{\Omega_{\rm{mw}}\Omega_1({\bf r})} {\Delta^2 -\Omega_{\rm{mw}}^2}\hat{\psi}_{00} - \frac{\Delta\Omega_2({\bf r})} {\Delta^2 -\Omega_{\rm{mw}}^2}\hat{\psi}_{1-1},
 \label{elim}
\end{align}
Substituting Eq.~(\ref{elim}) into Eq.~(\ref{dequs}), the dynamical equation of $\hat{\psi}_{NM_N}$ is given by
\begin{align}
i\dot{\hat \psi}_{00}
&=\frac{\Omega_{\rm{mw}} \Omega^*_1({\bf r})\Omega_2({\bf r})} {\Delta^2 -\Omega_{\rm{mw}}^2}\sum_{\sigma=\pm1}\hat{\psi}_{1\sigma} - \frac{2\Delta |\Omega_1({\bf r})|^2} {\Delta^2 -\Omega_{\rm{mw}}^2}\hat{\psi}_{00}, \nonumber \\
i\dot{\hat \psi}_{11} &= (\delta+\delta_{1,-1}){\hat \psi}_{11} + \frac{\Omega_{\rm{mw}}\Omega_1({\bf r})\Omega^*_2({\bf r})} {\Delta^2 -\Omega_{\rm{mw}}^2}\hat{\psi}_{00}
- \frac{\Delta|\Omega_2({\bf r})|^2} {\Delta^2 -\Omega_{\rm{mw}}^2}\hat{\psi}_{11}, \nonumber \\
i\dot{\hat \psi}_{1-1} &=  \delta{\hat \psi}_{1-1} + \frac{\Omega_{\rm{mw}}\Omega_1({\bf r})\Omega^*_2({\bf r})} {\Delta^2 -\Omega_{\rm{mw}}^2}\hat{\psi}_{00} - \frac{\Delta|\Omega_2({\bf r})|^2} {\Delta^2 -\Omega_{\rm{mw}}^2}\hat{\psi}_{1-1},
\end{align}
As we mentioned in the main text, we will denote the rotational states by the quantum number $M_N$ only for shorthand notation. As a result, one can derive an effective Hamiltonian for a pseudospin-1 molecule
\begin{align}
\hat {\cal H}=&\hbar(\delta+\delta_{1,-1}+V_2)\hat{\psi}^\dag_{1}\hat{\psi}_{1} + \hbar(\delta+V_2)\hat{\psi}^\dag_{-1}\hat{\psi}_{-1}+ 2\hbar V_1\hat{\psi}^\dag_{0}\hat{\psi}_{0}+\left[\hbar\Omega e^{i\kappa y} \sum_{\sigma=\pm1}\hat{\psi}^\dag_{\sigma}\hat{\psi}_{0} +{\rm H.c.}\right],
\end{align}
where $V_{1,2}={\Delta\Omega^2_{1,2}}/({\Delta^2-\Omega_{\rm{mw}}^2})$ are the optical Stark shifts. After taking into account the center-of-mass motion, the above Hamiltonian can be rewritten as (choosing $2\hbar V_1$ as the origin of the energies)
\begin{eqnarray}
\hat{\cal H}_{0}=\sum_{\sigma\sigma'}\int d{\mathbf r}\hat\psi_{\sigma}^{\dag}({\mathbf r})\left[\hat h_{\sigma\sigma'}+U({\mathbf r})\delta_{\sigma\sigma'}\right]\hat\psi_{\sigma}({\mathbf r}),
\end{eqnarray}
which corresponds to the effective single-particle Hamiltonian
\begin{align}
\hat h = \begin{pmatrix}
\frac{{\mathbf p}^2}{2M}+ \hbar\delta_1+\hbar\delta_2& \hbar\Omega e^{i\kappa y} & 0 \\
\hbar\Omega  e^{-i\kappa y}  & \frac{{\mathbf p}^2}{2M}  & \hbar\Omega e^{-i\kappa y}\\
0 &\hbar\Omega e^{i\kappa y}  & \frac{{\mathbf p}^2}{2M}-\hbar\delta_1 +\hbar\delta_2
 \end{pmatrix}.\label{ssing}
\end{align}
where $\delta_1 = \delta_{1,-1}/2$ is the Zeeman shift and $\delta_2 = \delta+V_2 + \delta_{1,-1}/2-2V_1$ is the quadratic Zeeman shift for spin-1 polar molecules. Finally, we obtain the effective spin-1 single-molecule Hamiltonian (1) in the main text.

\end{widetext}


\begin{thebibliography}{99}

\bibitem{KRb-exp} K.-K. Ni, S. Ospelkaus, M.H.G. de Miranda, A. P\'{e}er, B. Neyenhuis, J.J. Zirbel, S. Kotochigova, P.S. Julienne, D.S. Jin, and J. Ye, Science {\bf 322}, 231 (2008).

\bibitem{LiCs-exp} J. Deiglmayr, A. Grochola, M. Repp, K. M\"{o}rtlbauer, C. Gl\"{u}ck, J. Lange, O. Dulieu, R. Wester, and M. Weidem\"{u}ller, Phys. Rev. Lett. {\bf101}, 133004 (2008).

\bibitem{RbCs-exp} T. Takekoshi, L. Reichs\"{o}llner, A. Schindewolf, J.M. Hutson, C. R. Le Sueur, O. Dulieu, F. Ferlaino, R. Grimm, and H.-C. N\"{a}gerl, Phys. Rev. Lett. {\bf113}, 205301 (2014).

\bibitem{RbCs-exp2} T. Shimasaki, M. Bellos, C.D. Bruzewicz, Z. Lasner, and D. DeMille, Phys. Rev. A {\bf91}, 021401(R) (2015).

\bibitem{RbCs-exp3} P.K. Molony,  P.D. Gregory, Z. Ji, B. Lu, M.P. K\"{o}ppinger, C. R. Le Sueur, C.L. Blackley, J.M. Hutson, and S.L. Cornish, Phys. Rev. Lett. {\bf113}, 255301 (2014).

\bibitem{NaK-exp} J.W. Park, S.A. Will, and M.W. Zwierlein, Phys. Rev. Lett. {\bf114}, 205302 (2015).

\bibitem{Guo16} M. Guo, B. Zhu, B. Lu, X. Ye, F. Wang, R. Vexiau, N. Bouloufa-Maafa, G. Qu\'{e}m\'{e}ner, O. Dulieu, and D. Wang, Phys. Rev. Lett. {\bf116}, 205303 (2016).

\bibitem{Rvachov17} T.M. Rvachov, H. Son, A.T. Sommer, S. Ebadi, J.J. Park, M.W. Zwierlein, W. Ketterle, and A.O. Jamison, Phys. Rev. Lett. {\bf119}, 143001 (2017).

\bibitem{krb-coll} K.-K. Ni, S. Ospelkaus, D. Wang, G. Qu\'{e}m\'{e}ner, B. Neyenhuis, M.H.G. de Miranda, J.L. Bohn, J. Ye, and D.S. Jin, Nature (London) {\bf464}, 1324 (2010).

\bibitem{krb-chem} S. Ospelkaus, K.-K. Ni, D. Wang, M.H.G. de Miranda, B. Neyenhuis, G. Qu\'{e}m\'{e}ner, P.S. Julienne, J.L. Bohn, D.S. Jin, and J. Ye, Science {\bf327}, 853 (2010).

\bibitem{Miranda11} M.H.G. de Miranda, A. Chotia1, B. Neyenhuis, D.Wang, G. Qu\'{e}m\'{e}ner, S. Ospelkaus, J.L. Bohn, J. Ye, and D.S. Jin, Nat. Phys. {\bf 7}, 502 (2011).

\bibitem{qucompu} D. DeMille, Phys. Rev. Lett. {\bf88}, 067901  (2002).

\bibitem{qu-info} P. Rabl, D. DeMille, J.M. Doyle, M.D. Lukin, R.J. Schoelkopf, and P. Zoller, Phys. Rev. Lett. {\bf97}, 033003 (2006).

\bibitem{Micheli06} A. Micheli, G.K. Brennen, and P. Zoller, Nat. Phys. {\bf2}, 341 (2006).

\bibitem{Carr09} L.D. Carr, D. DeMille, R.V. Krems, and J. Ye, New J. Phys. {\bf11}, 055049 (2009).

\bibitem{pre-measu1} V.V. Flambaum and M.G. Kozlov, Phys. Rev. Lett. {\bf99}, 150801 (2007).

\bibitem{pre-measu2} T.A. Isaev, S. Hoekstra, and R. Berger, Phys. Rev. A {\bf82}, 052521 (2010).

\bibitem{pre-measu3} J.J. Hudson, D.M. Kara, I.J. Smallman, B.E. Sauer, M.R. Tarbutt, and E.A. Hinds, Nature (London) {\bf473}, 493 (2011).

\bibitem{Cairncross17} W.B. Cairncross, D.N. Gresh, M. Grau, K.C. Cossel, T.S. Roussy, Y. Ni, Y. Zhou, J. Ye, E.A. Cornell, Phys. Rev. Lett. {\bf119}, 153001 (2017).

\bibitem{Pasquiou11}  B. Pasquiou, E. Mar\'{e}chal, G. Bismut, P. Pedri, L. Vernac, O. Gorceix, and B. Laburthe-Tolra, Phys. Rev. Lett. {\bf106}, 255303 (2011).

\bibitem{Aikawa14} K. Aikawa, S. Baier, A. Frisch, M. Mark, C. Ravensbergen, and F. Ferlaino, Science {\bf345}, 1484 (2014).

\bibitem{Kadau16} H. Kadau, M. Schmitt, M. Wenzel, C. Wink, T. Maier, I. Ferrier-Barbut, and T. Pfau, Nature (London) {\bf530}, 194 (2016).

\bibitem{Schmitt16} M. Schmitt, M. Wenzel, F. B\"{o}ttcher, I. Ferrier-Barbut, and T. Pfau, Nature (London) {\bf539}, 259 (2016).

 \bibitem{Barbut16} I. Ferrier-Barbut, H. Kadau, M. Schmitt, M. Wenzel, and T. Pfau, Phys. Rev. Lett. {\bf116}, 215301 (2016).

\bibitem{Lin11} Y.-J. Lin, K. Jim\'{e}nez-Garc\'{i}a, and I.B. Spielman, Nature (London) {\bf 471}, 83 (2011).

\bibitem{Zhang12} J.-Y. Zhang, S.-C. Ji, Z. Chen, L. Zhang, Z.-D. Du, B. Yan, G.-S. Pan, B. Zhao, Y.-J. Deng, H. Zhai, S. Chen, and J.-W. Pan, Phys. Rev. Lett. {\bf 109}, 115301 (2012).

\bibitem{Qu13} C. Qu, C. Hamner, M. Gong, C. Zhang, and P. Engels, Phys. Rev. A {\bf 88}, 021604(R) (2013).

\bibitem{Wu15} Z. Wu, L. Zhang, W. Sun, X.-T. Xu, B.-Z. Wang, S.-C. Ji, Y. Deng, S. Chen, X.-J. Liu, and J.-W. Pan, Science {\bf354}, 83 (2016).

\bibitem{Campbell16} D.L. Campbell, R.M. Price, A. Putra, A. Vald\'{e}s-Curiel, D. Trypogeorgos, and I.B. Spielman, Nat. Commun {\bf7}, 10897 (2016).

\bibitem{Luo16} X. Luo, L. Wu, J. Chen, Q. Guan, K. Gao, Z.-F. Xu, L. You, and R. Wang, Sci. Rep. {\bf6}, 18983 (2016).

\bibitem{Li17} J.-R. Li, J. Lee, W. Huang, S. Burchesky, B. Shteynas, F.C. Top, A.O. Jamison, and W. Ketterle, Nature (London) {\bf 543}, 91 (2017).

\bibitem{Wang12} P. Wang, Z.-Q. Yu, Z. Fu, J. Miao, L. Huang, S. Chai, H. Zhai, and J. Zhang, Phys. Rev. Lett. {\bf 109}, 095301 (2012).

\bibitem{Cheuk12} L.W. Cheuk, A.T. Sommer, Z. Hadzibabic, T. Yefsah, W.S. Bakr, and M.W. Zwierlein, Phys. Rev. Lett. {\bf 109}, 095302 (2012).

\bibitem{Huang15} L. Huang, Z. Meng, P. Wang, P. Peng, S.-L. Zhang, L. Chen, D. Li, Q. Zhou, and J. Zhang, Nat. Phys. {\bf12}, 540 (2016).

\bibitem{Burdick16} N.Q. Burdick, Y. Tang, and B.L. Lev, Phys. Rev. X {\bf 6}, 031022 (2016).

\bibitem{Song16} B. Song, C. He, S. Zhang, E. Hajiyev, W. Huang, X.-J. Liu, and G.-B. Jo, Phys. Rev. A {\bf 94}, 061604(R) (2016).

\bibitem{Deng2012} Y. Deng, J. Cheng, H. Jing, C.-P. Sun, and S. Yi, Phys. Rev. Lett. {\bf 108}, 125301 (2012).

\bibitem{Wilson2013} R.M. Wilson, B.M. Anderson, and C.W. Clark, Phys. Rev. Lett. {\bf 111}, 185303 (2013).

\bibitem{Ng2014} H.T. Ng, Phys. Rev. A {\bf 90}, 053625 (2014).

\bibitem{Deng2015} Y. Deng and S. Yi, Phys. Rev. A {\bf 92}, 033624 (2015).

\bibitem{Wall15} M.L. Wall, K. Maeda, and L.D. Carr, New J. Phys. {\bf17}, 025001 (2015).

\bibitem{Lan2014} Z. Lan and P. \"{O}hberg, Phys. Rev. A {\bf89}, 023630 (2014).

\bibitem{Natu2015} S.S. Natu, X. Li, and W.S. Cole, Phys. Rev. A {\bf91}, 023608 (2015).

\bibitem{Luo17} X.-W. Luo, K. Sun, and C. Zhang, Phys. Rev. Lett. {\bf 119}, 193001 (2017).

\bibitem{youxu} Z. F. Xu, Y. Zhang, and L. You, Phys. Rev. A {\bf79}, 023613 (2009); Z. F. Xu, J. Zhang, Y. Zhang, and L. You, {\it ibid} {\bf81}, 033603 (2010).

\bibitem{Ueda2007} Y. Kawaguchi, H. Saito, and M. Ueda, Phys. Rev. Lett. {\bf98}, 110406 (2007).

\bibitem{Sonin87} E.B. Sonin, Rev. Mod. Phys. {\bf59}, 87 (1987).

\bibitem{Kopnin02} N.B. Kopnin, Rep. Prog. Phys. {\bf65}, 1633 (2002).

\bibitem{Bloch08} I. Bloch, J. Dalibard, and W. Zwerger, Rev. Mod. Phys. {\bf80}, 885 (2008).

\bibitem{Fetter09} A.L. Fetter, Rev. Mod. Phys. {\bf81}, 647 (2009).

\bibitem{atom-dipspin-theo1} S. Yi and H. Pu, Phys. Rev. Lett. {\bf97}, 020401 (2006).

\bibitem{Yi2007} S. Yi, T. Li, and C.P. Sun, Phys. Rev. Lett. {\bf98}, 260405 (2007).

\bibitem{atom-dipspin-exp1} M. Vengalattore, S.R. Leslie, J. Guzman, and D.M. Stamper-Kurn, Phys. Rev. Lett. {\bf100}, 170403 (2008).

\bibitem{Yao2012} N.Y. Yao, C.R. Laumann, A.V. Gorshkov, S.D. Bennett, E. Demler, P. Zoller, and M.D. Lukin, Phys. Rev. Lett. {\bf109}, 266804 (2012).

\bibitem{Yao2013} N.Y. Yao, A.V. Gorshkov, C.R. Laumann, A.M. L\"{a}uchli, J. Ye, and M.D. Lukin, Phys. Rev. Lett. {\bf110}, 185302 (2013).

\bibitem{Peter2015} D. Peter, N.Y. Yao, N. Lang, S.D. Huber, M.D. Lukin, and H.P. B\"{u}chler, Phys. Rev. A {\bf91}, 053617 (2015).

\bibitem{Syzranov2015} S.V. Syzranov, M.L. Wall, B. Zhu, V. Gurarie, and A.M. Rey, Nat. Commun. {\bf7}, 243002 (2016).

\end{thebibliography}
\end{document}